\documentclass[showpacs,doublespace,preprintnumbers,amsmath,amssymb,prb]{revtex4}

\usepackage{bm}

\begin{document}

\begin{large}
\title{Critical behavior of long straight rigid rods on
two-dimensional lattices: Theory and Monte Carlo simulations}

\author{D. A. Matoz-Fernandez}
\affiliation{Departamento de F\'{\i}sica, Instituto de F\'{\i}sica
Aplicada, Universidad Nacional de San Luis-CONICET, Chacabuco 917,
D5700BWS San Luis, Argentina}
\author{D. H. Linares}
\affiliation{Departamento de F\'{\i}sica, Instituto de F\'{\i}sica
Aplicada, Universidad Nacional de San Luis-CONICET, Chacabuco 917,
D5700BWS San Luis, Argentina}
\author{A.J. Ramirez-Pastor$^{\dag}$}
\affiliation{Departamento de F\'{\i}sica, Instituto de F\'{\i}sica
Aplicada, Universidad Nacional de San Luis-CONICET, Chacabuco 917,
D5700BWS San Luis, Argentina}

\date{\today}

\begin{abstract}

The critical behavior of long straight rigid rods of length $k$
($k$-mers) on square and triangular lattices at intermediate
density has been studied. A nematic phase, characterized by a big
domain of parallel $k$-mers, was found. This ordered phase is
separated from the isotropic state by a continuous transition
occurring at a intermediate density $\theta_c$. Two analytical
techniques were combined with Monte Carlo simulations to predict
the dependence of $\theta_c$ on $k$, being $\theta_c(k) \propto
k^{-1}$. The first involves simple geometrical arguments, while
the second is based on entropy considerations. Our analysis
allowed us also to determine the minimum value of $k$
($k_{min}=7$), which allows the formation of a nematic phase on a
triangular lattice.
\end{abstract}


\maketitle

\noindent $\dag$ To whom all correspondence should be addressed.

\newpage

\section{Introduction}

The isotropic-nematic (I-N) phase transition has been a topic of
active theoretical and experimental studies over the past few
decades. An early seminal contribution to this subject was made by
Onsager~\cite{ONSAGER} with his work on the I-N phase transition
of infinitely thin rods. This theory predicted that excluded
volume interactions alone can lead to long-range orientational
(nematic) order. Later, several papers contributed greatly to the
understanding of the statistics of rigid
rods~\cite{FLORY,HUGGI,GUGGE,DIMA,ZWANZIG,FREED}. Successive works
have established detailed phase diagrams for several hard-body
shapes~\cite{LEE,SAMBOR,MARGO,ALLEN}. For the continuum problem,
there is general agreement that in three dimensions, infinitely
thin rods undergo a first-order I-N transition, as was pointed out
by Onsager~\cite{ONSAGER}. In two dimensions, the nature of the
isotropic-nematic (I-N) phase transition depends crucially on the
particle interactions and a rich variety of behaviors is
observed~\cite{STRALEY,FRENKEL}.

A notable feature is that nematic order is only stable for
sufficiently large aspect ratios while isotropic systems of short
rods not show nematic order at all. The long-range orientational
order also disappears in the case of irreversible adsorption (no
desorption)~\cite{HINRICHSEN,EVANS}, where the distribution of
adsorbed objects is different from that obtained at
equilibrium~\cite{TALBOT,KONDRAT}. Thus, at high coverage, the
equilibrium state corresponds to a nematic phase with long-range
correlations, whereas the final state generated by irreversible
adsorption has infinite memory of the process and orientational
order is purely local.

In the case of lattice models, two previous
articles~\cite{MATOZ,LINARES}, referred to as papers I and II,
respectively, were devoted to the study of the I-N phase
transition in a system of long straight rigid rods of length $k$
($k$-mers) on two-dimensional lattices with discrete allowed
orientations. The model of a two-dimensional gas of rigid $k$-mers
is the simplest representation of a strongly adsorbed film of
linear molecules in submonolayer or monolayer regime. Examples of
this kind of systems are monolayer films of $n$-alkanes adsorbed
on monocrystalline surfaces of metals, such as
Pt(111)~\cite{HOSTETLER} and Au(111)~\cite{POTOFF1,POTOFF2}.

Papers I and II were inspired in the excellent paper by Ghosh and
Dhar~\cite{GHOSH}. In Ref.~[\onlinecite{GHOSH}], the authors
presented strong numerical evidence that a system of square
geometry, with two allowed orientations, shows nematic order at
intermediate densities for $k \geq 7$ and provided a qualitative
description of a second phase transition (from a nematic order to
a non-nematic state) occurring at a density close to $1$. However,
the authors were not able to determine the critical quantities
(critical point and critical exponents) characterizing the I-N
phase transition occurring in the system.

In this context, extensive Monte Carlo (MC) simulations were used
in paper I to resolve the universality class of the first phase
transition occurring in the system of Ref.~[\onlinecite{GHOSH}].
Lattices of various sizes were considered and finite-size scaling
theory was utilized. As it was evident from the calculation of the
critical exponents and the behavior of Binder cumulants, the
universality class was shown to be that of the 2D Ising model for
square lattices with two allowed orientations, and the three-state
Potts model for triangular lattices with three allowed
orientations.

Paper II was a step further, analyzing the I-N phase transition in
terms of entropy. For this purpose, the configurational entropy of
a system of rigid rods deposited on a square lattice was
calculated by Monte Carlo (MC) simulations and thermodynamic
integration method~\cite{BINDER1}. The numerical data were
compared with the corresponding ones obtained from a fully aligned
system (nematic phase), whose calculation reduces to the
one-dimensional case~\cite{PRB3}. The study allowed us $1)$ to
confirm previous results in the literature~\cite{GHOSH}, namely,
the existence of $i)$ a I-N phase transition at intermediate
densities for $k \geq 7$ and $ii)$ a second phase transition from
a nematic order to a non-nematic state at high density. In the
second case, Ref.~[\onlinecite{LINARES}] represents the first
numerical evidence existing in the literature about this important
point; $2)$ to provide an interpretation on the underlying physics
of the observed $k$ dependence of the isotropic-nematic phase
transition; and $3)$ to test the predictions of the main
theoretical models developed to study adsorption with
multisite-occupancy.

Even though many aspects of the problem have been treated in
papers I, II and Ref.~[\onlinecite{GHOSH}], other points remain
still open. Among them, there exists no studies on the dependence
of the critical density characterizing the I-N phase transition,
$\theta_c$, on the size $k$ of the rod. In this work we attempt to
remedy this situation. For this purpose, extensive MC simulations
complemented by finite-size scaling techniques and theoretical
modeling have been applied. We restrict the study to the first
phase transition occurring in the system (or I-N phase transition
at intermediate density). Our analysis allowed us $(1)$ to obtain
$\theta_c$ as a function of $k$ for square and triangular
lattices, being $\theta_c(k) \propto k^{-1}$; and $(2)$ to
determine the minimum value of $k$ ($k_{min}=7$), which allows the
formation of a nematic phase on triangular lattices.

The outline of the paper is as follows. In Sec. II we describe the
lattice-gas model, the simulation scheme, and we present the
behavior of $\theta_c(k)$, obtained by using the MC method. In
Sec. III we present the analytical approximations and compare the
MC results with the theoretical calculations. Finally, the general
conclusions are given in Sec. IV.

\section{Lattice-gas model and Monte Carlo simulation scheme}
\subsection{Model and Monte Carlo method}

We address the general case of adsorbates assumed to be linear
rigid particles containing $k$ identical units ($k$-mers), with
each one occupying a lattice site. Small adsorbates would
correspond to the monomer limit ($k = 1$). The distance between
$k$-mer units is assumed to be equal to the lattice constant;
hence exactly $k$ sites are occupied by a $k$-mer when adsorbed.
The only interaction between different rods is hard-core
exclusion: no site can be occupied by more than one $k$-mer unit.
The surface is represented as an array of $M = L \times L$
adsorptive sites in a square or triangular lattice arrangement,
where $L$ denotes the linear size of the array.

The degree of order in the adsorbed phase is calculated for each
configuration according to the standard method used for the Potts
model~\cite{WU}. To this end, we first build a set of vectors
$\{\vec{n}_1,\vec{n}_2,\cdots,\vec{n}_m\}$ with the following
properties: $(i)$ each vector is associated to one of the $m$
possible orientations (or directions) for a $k$-mer on the
lattice; $(ii)$ the $\vec{n}_i$'s lie in the same plane (or are
co-planar) and point radially outward from a given point $P$ which
is defined as coordinate origin; $(iii)$ the angle between two
consecutive vectors, $\vec{n}_i$ and $\vec{n}_{i+1}$, is equal to
$2\pi/m$; and $(iv)$ the magnitude of $\vec{n}_i$ is equal to the
number of $k$-mers aligned along the $i$-direction. Note that the
$\vec{n}_i$'s have the same directions as the $q$ vectors in
Ref.~\cite{WU}. These directions are not coincident with the
allowed directions for the $k$-mers on the real lattice. Then the
order parameter $\delta$ of the system is given by
\begin{equation}
\delta =  \frac{\left | \sum_{i=1}^m \vec{n}_i \right
|}{\sum_{i=1}^m \left | \vec{n}_i \right |}
 \label{fi}
\end{equation}
$\delta$ represents a general order parameter measuring the
orientation of the $k$-mers on a lattice with $m$ directions. In
the case of square lattices, $m=2$ and the angle between
$\vec{n}_1$ and $\vec{n}_{2}$ is $\pi$. Accordingly, the order
parameter reduces to $\delta =  \left | n_1 -n_2 \right |/ \left (
n_1+n_2 \right )$, $n_1$ ($n_2$) being the number of $k$-mers
aligned along the horizontal (vertical) direction. This expression
coincides with the order parameter $Q$ defined in
Ref.~[\onlinecite{GHOSH}]. On the other hand, $m=3$ for triangular
lattices and $\vec{n}_1$, $\vec{n}_{2}$ and $\vec{n}_{3}$ form
angles of $2\pi/3$ between them.

When the system is disordered $(\theta<\theta_c)$, all
orientations are equivalents and $\delta$ is zero. As the density
is increased above $\theta_c$, the $k$-mers align along one
direction and $\delta$ is different from zero. Thus, $\delta$
appears as a proper order parameter to elucidate the phase
transition.

The problem has been studied by grand canonical MC simulations
using a typical adsorption-desorption
algorithm~\cite{LANG7,LANG10,SS12}. The procedure is as follows.
Once the value of the chemical potential $\mu$ is set, a linear
$k$-uple of nearest-neighbor sites is chosen at random. Then, if
the $k$ sites are empty, an attempt is made to deposit a rod with
probability $W={\rm min} \left\{1,\exp\left( \mu/k_BT \right)
\right\}$, where $k_B$ is the Boltzmann constant and $T$ is the
temperature; if the $k$ sites are occupied by units belonging to
the same $k$-mer, an attempt is made to desorb this $k$-mer with
probability $W={\rm min} \left\{1,\exp\left( -\mu/k_BT \right)
\right\}$; and otherwise, the attempt is rejected. In addition,
displacement (diffusional relaxation) of adparticles to
nearest-neighbor positions, by either jumps along the $k$-mer axis
or reptation by rotation around the $k$-mer end, must be allowed
in order to reach equilibrium in a reasonable time. A MC step
(MCs) is achieved when $M$ $k$-uples of sites have been tested to
change its occupancy state. Typically, the equilibrium state can
be well reproduced after discarding the first $r'=10^6$ MCs. Then,
the next $r=2 \times 10^6$ MCs are used to compute averages.

In our Monte Carlo simulations, we varied the chemical potential
$\mu$ and monitored the density $\theta$ and the order parameter
$\delta$, which can be calculated as simple averages. The reduced
fourth-order cumulant $U_L$ introduced by Binder~\cite{BINDER} was
calculated as:
\begin{equation}
U_L = 1 -\frac{\langle \delta^4\rangle} {3\langle
\delta^2\rangle^2}, \label{ul}
\end{equation}
where $\langle \cdots \rangle$ means the average over the $r$ MC
simulation runs.

Finally, the configurational entropy of the system $S$, was
calculated by using thermodynamic integration method
~\cite{HANSEN,BINDER2,BINDER3,POLGREEN}. The method in the grand
canonical ensemble relies upon integration of the chemical
potential $\mu$ on coverage along a reversible path between an
arbitrary reference state and the desired state of the system.
This calculation also requires the knowledge of the total energy
$U$ for each obtained coverage. Thus, for a system made of $N$
particles on $M$ lattice sites, we have:
\begin{eqnarray}
S(N,M,T) & = & S(N_0,M,T)+{U(N,M,T)-U(N_0,M,T) \over T}
\nonumber\\
& &- {1\over T} \int_{N_0}^N{ \mu dN'}\label{entN}
\end{eqnarray}
In our case $U(N,M,T)=0$ and the determination of the entropy in
the reference state, $S(N_0,M,T)$, is trivial [$S(N_0,M,T)=0$ for
$N_0 = 0$]. Note that the reference state, $N \rightarrow 0$, is
obtained for $\mu/k_B T \rightarrow -\infty$. Then,
\begin{equation}
{s(\theta,T) \over k_B}= - {1 \over k_B T} \int_{0}^{\theta}{
\frac{\mu}{k} ~ d \theta'}\label{ent}
\end{equation}
where $s(=S/M)$ is the configurational entropy per site and
$\theta(=k~N/M)$ is the surface coverage (or density).

\subsection{Computational results}

Computational simulations have been developed for a system of
straight rigid rods of length $k$ ($k=2-14$) on a lattice. The
surface was represented as an array of adsorptive sites in a
square or triangular lattice arrangement. In addition,
conventional periodic boundary conditions were considered. The
effect of finite size was investigated by examining square
lattices with $L/k=5, 10, 15, 20$ and triangular lattices with
$L/k=10, 15, 20, 25$.

As it was established in Ref.~[\onlinecite{GHOSH}], the minimum
value of $k$, which allows the formation of a nematic phase on a
square lattice at intermediate densities, is $k_{min}=7$. This
critical quantity has not been calculated yet for other
geometries. Then, our first objective is to obtain $k_{min}$ for
triangular lattices. For this purpose, two criteria have been
applied: $1)$ the comparison between the configurational entropy
of the system, obtained by MC simulations, and the corresponding
to a fully aligned system (nematic phase), whose calculation
reduces to the 1D case~\cite{LINARES}; and $2)$ the behavior of
the nematic order parameter $\delta$ as a function of coverage.

The results obtained in the first case are shown in Fig. 1. Dotted
line and symbols represent MC data for triangular lattices,
$k=6,7$ and $8$  and $L/k=20$. The calculation of $s(\theta)/k_B$
through eq.~(\ref{ent}) is straightforward and computationally
simple, since the coverage dependence of $\mu/k_BT$ is evaluated
following the standard procedure of MC simulation described in
previous section. Then, $\mu(\theta)/k_BT$ is spline-fitted and
numerically integrated.

On the other hand, when the nematic phase is formed, the system is
characterized by a big domain of parallel $k$-mers. The
calculation of the entropy of this fully aligned state having
density $\theta$ reduces to the calculation of a one-dimensional
problem~\cite{PRB3}
\begin{eqnarray}
{s(\theta) \over k_B} & = &  \left[1-{\left(k-1 \right) \over k}
\theta \right]\ln \left[1-{\left(k-1 \right) \over k} \theta
\right] \nonumber\\
& & - {\theta \over k}\ln{\theta \over k} -\left(1-\theta
\right)\ln \left(1-\theta \right). \label{s1d}
\end{eqnarray}
Results from eq.~(\ref{s1d}) for $k=6,7$ and $8$ are shown in Fig.
1 (solid lines). As it can be observed, for $k \leq 6$, the 1D
results present a smaller $s/k_B$ than the 2D simulation data over
all the range of $\theta$. For $k \geq 7$, there exists a range of
coverage for which the difference between the 1D value and the
true 2D value is very small. In other words, for $k \geq 7$ and
intermediate densities, it is more favorable for the rods to align
spontaneously, since the resulting loss of orientational entropy
is by far compensated by the gain of translational entropy. This
finding is a clear indication that $k_{min}=7$ for triangular
lattices. In addition, $\theta_c$ can be calculated from the
minimum value of $\theta$ for which the near superposition of the
1D and 2D results occurs. Thus, the technique provides an
alternative method of determining the critical coverage
characterizing the I-N phase transition without any special
requirement and time consuming computation. However, it is
important to emphasize that the calculation of the entropy of the
nematic phase from the 1D model is an approximation (especially at
the moderate densities, where the phase is not completely
aligned). Consequently, a precise determination of $\theta_c$
should require an extensive work of MC simulation and finite-size
scaling techniques.

As an independent corroboration of the results previously
obtained, the inset in Fig. 1 presents the nematic order parameter
$\delta$ as a function of coverage for $k=6,7$ and $8$ and
$L/k=20$. The behavior of $\delta$ confirms the existence of
nematic order for $k \geq 7$. This value of $k_{min}$ coincides
with that obtained for square lattices.

Once $k_{min}$ has been established, we now discuss the behavior
of the critical density as a function of the size $k$. In the case
of the standard theory of finite-size
scaling~\cite{BINDER,PRIVMAN}, when the phase transition is
temperature driven, the technique allows for various efficient
routes to estimate $T_c$ from MC data. One of these methods, which
will be used in this case, is from the temperature dependence of
$U_L(T)$, which is independent of the system size for $T=T_c$. In
other words, $T_c$ is found from the intersection of the curve
$U_L(T)$ for different values of $L$, since $U_L(T_c)=const$. In
our study, we modified the conventional finite-size scaling
analysis by replacing temperature by density~\cite{MATOZ}. Under
this condition, the critical density has been estimated from the
plots of the reduced four-order cumulants $U_L(\theta)$ plotted
versus $\theta$ for several lattice sizes. As an example, Fig. 2
shows the results for rods of size $k=10$ on square [Fig. 2(a)]
and triangular [Fig. 2(b)] lattices. In the cases of the figure,
the values obtained were $\theta_c=0.502(1)$ and
$\theta_c=0.530(1)$, for square and triangular lattices,
respectively. The curves of the order parameter, which were used
to obtain $U_L(\theta)$, are shown in the insets of the figure.

The procedure of Fig. 2 was repeated for $7 \leq k \leq 14$ and
the results are collected in Fig. 3. The log-log plots show that
the critical density follows a power law as $\theta_c(k) \propto
k^{-1}$. The understanding of this dependence of $\theta_c(k)$ on
$k$ can be developed by using simple geometric arguments. An
example follows in order to make this point clear. Fig. 4(a) shows
a snapshot corresponding to a ``small window" of side $l$, which
is embedded in an infinite square lattice. Open squares correspond
to empty sites and black bars represent adsorbed $k$-mers (in the
case of the figure, $k=4$ and $l=8$). Then, the local coverage or
fraction of occupied sites belonging to the window is
$\theta=kN/(l \times l)$, being $N$ the number of $k$-mers into
the window~\cite{FOOT1}. We can now think of a transformation $L
\rightarrow L'$ from the original lattice $L$ to a new lattice
$L'$, where each characteristic length of $L'$ is $s$ times the
corresponding one of $L$ ($s= 1,2,3,4, \cdots$). Then, $l'=sl$,
$k'=sk$ and $\theta'=k'N/(l' \times l')$. By following this
procedure, Fig. 4(b) was obtained from Fig. 4(a). In this case,
$s=3$, and consequently, $l'=3l=24$ and $k'=3k=12$. As it can be
observed, $L$ and $L'$ represent the ``same situation" for two
different values of $k$.

In general,
\begin{equation}
{\theta' \over \theta}  = {k'N (l \times l) \over k N (l' \times
l')} \label{t1ont2},
\end{equation}
and $l/k=l'/k'$. Then, the relationship between a coverage in $L$
and the corresponding one in $L'$ results
\begin{equation}
{\theta \over \theta'}  = {k' \over k} \label{t1ont2a}.
\end{equation}
 By taking the system
$L'$ as reference, it is possible to write,
\begin{eqnarray}
{\theta }  & = & (k' \theta') k^{-1} \nonumber\\
& \propto & k^{-1} \label{t1ont2a},
\end{eqnarray}
as it was found from the MC data. A more rigorous justification of
the observed $k$ dependence of the I-N phase transition will be
presented in the next section.

It is worth pointing out that we did not assume any particular
universality class for the transitions analyzed here in order to
calculate their critical temperatures, since the analysis relied
on the order parameter cumulant's properties. However, a
systematic analysis of critical exponents was carried out in
Ref.~[\onlinecite{MATOZ}], where the universality class was shown
to be that of the 2D Ising model for square lattices and the
two-dimensional Potts model with $q=3$ for triangular lattices.

\section{Analytical approximations and comparison between
 simulated and theoretical results}

Next, the transition is analyzed from the main theoretical models
developed to treat the polymers adsorption problem. The study
allows us to calculate the dependence of $\theta_c(k)$ on $k$ from
three  well-known multisite-adsorption theories: the Flory-Huggins
(FH) approximation~\cite{FLORY,HUGGI}, the Guggenheim-DiMarzio
(GD) approach~\cite{GUGGE,DIMA} and the recently developed
Semiempirical Model for Adsorption of Polyatomics
(SE)~\cite{LANG11,IJMP}. The corresponding expressions for the
configurational entropy per site are:
\begin{equation}
{s(\theta) \over k_B} =  - {\theta \over k}\ln{\theta \over k}
-\left(1-\theta \right)\ln \left(1-\theta \right) - {\theta \over
k} \left[k-1- \ln \left( c \over 2 \right) \right] \  \  \  \  \ \
\ {\rm (FH) \label{fh}}.
\end{equation}
\begin{equation}
{s(\theta) \over k_B}  =   - {\theta \over k}\ln{\theta \over k}
-\left(1-\theta \right)\ln \left(1-\theta \right) + \left(\theta-
{c \over 2} \right) \ln \left( c \over 2 \right)  + \left[{c \over
2}-{\left(k-1 \right) \over k} \theta \right]\ln \left[{c \over
2}-{\left(k-1 \right) \over k} \theta \right] \ \ \ \ \  \ \ \ \
{\rm (GD) \label{gd}}.
\end{equation}
and
\begin{eqnarray}
{s(\theta) \over k_B} & = &  - {\theta \over k}\ln{\theta \over k}
-\left(1-\theta \right)\ln \left(1-\theta \right) \nonumber\\
& & +
\theta \left[{1 \over 2} - {c \over 4} + {1 \over k} \ln \left( c
\over 2 \right) \right] +{1 \over 2} {k \over \left(k-1 \right)}
\left[1-{\left(k-1 \right)^2 \over k^2} \theta^2 \right]\ln
\left[1-{\left(k-1
\right) \over k} \theta \right] \nonumber\\
& & - {c \over 4} \left[\theta + {k \left(c-4 \right)+4 \over 2
\left(k-1 \right)} \right] \left[1-{2 \left(k-1 \right) \over c k}
\theta \right]\ln \left[1-{2 \left(k-1 \right) \over c k} \theta
\right] \ \ \ \ \ \ \  {\rm (SE) \label{se}}.
\end{eqnarray}

From a given value of $k$ (which depends on the approximation
considered), the 2D [eqs.~(\ref{fh}-\ref{se})] and 1D
[eq.~(\ref{s1d})] curves cross at intermediate densities and two
well differentiated regimes can be observed. In the first regime,
which occurs at low densities, the 2D approaches predict a larger
entropy than the 1D data. In the second regime (at high densities)
the behavior is inverted and the 2D data present a smaller $s/k_B$
than the 1D results. Given that the theoretical results in 2D
assume isotropy in the adlayer (interested readers are referred to
Ref.~[\onlinecite{DIMA}], where this point is explicitly
considered), the crossing of the curves shows that, in the second
regime, the contribution to the 2D entropy from the isotropic
configurations is lower than the contribution from the aligned
states. Then, the existence of an intersection point is indicative
of a I-N transition and allows us to estimate $k_{min}$ and
$\theta_c$ from the different approximations studied. This
intersection point can be easily calculated through a standard
computing procedure; in our case, we used Mathematica software.
The results are shown in Fig. 5 for the different studied cases.
The symbology is indicated in the figure.

In the case of square lattices, FH and GD predict values of
$k_{min}=3$ and $k_{min}=4$, respectively. On the other hand, SE
performs significantly better than the other approaches,
predicting the ``exact" value of $k_{min}=7$~\cite{GHOSH}. With
respect to triangular lattices, the values obtained for the
critical $k$ are $k_{min}=4$ (FH), $k_{min}=4$ (GD) and
$k_{min}=10$ (SE). Finally, the behavior of $\theta_c$ as a
function of $k$ for the analytical approximations follows
qualitatively the Monte Carlo simulation results, which reinforces
the robustness of the analysis introduced here.

\section{Conclusions}

In the present work, we have addressed the critical properties of
long straight rigid rods on square and triangular lattices at
intermediate density. The results were obtained by combining Monte
Carlo simulations, finite-size scaling techniques and three
well-known multisite-adsorption theories: the Flory-Huggins
approximation, the Guggenheim-DiMarzio approach and the
Semiempirical Model for Adsorption of Polyatomics.

Two main conclusions can be drawn from the present work. On one
hand, the critical density dependence on the particle size $k$ has
been reported. We found that $\theta_c(k)$ follows a power law as
$\theta_c(k) \propto k^{-1}$. On the other hand, our analysis
allowed us also to determine the minimum value of $k$
($k_{min}=7$), which allows the formation of a nematic phase on a
triangular lattice.

\acknowledgments This work was supported in part by CONICET
(Argentina) under project PIP 6294;  Universidad Nacional de San
Luis (Argentina) under project 322000 and the National Agency of
Scientific and Technological Promotion (Argentina) under project
33328 PICT 2005. The numerical work were done using the BACO
parallel cluster (composed by  60 PCs each with a 3.0 GHz
Pentium-4 processor) located  at Instituto de F\'{\i}sica
Aplicada, Universidad Nacional de San Luis-CONICET, San Luis,
Argentina.

\newpage
\section*{Figure Captions}

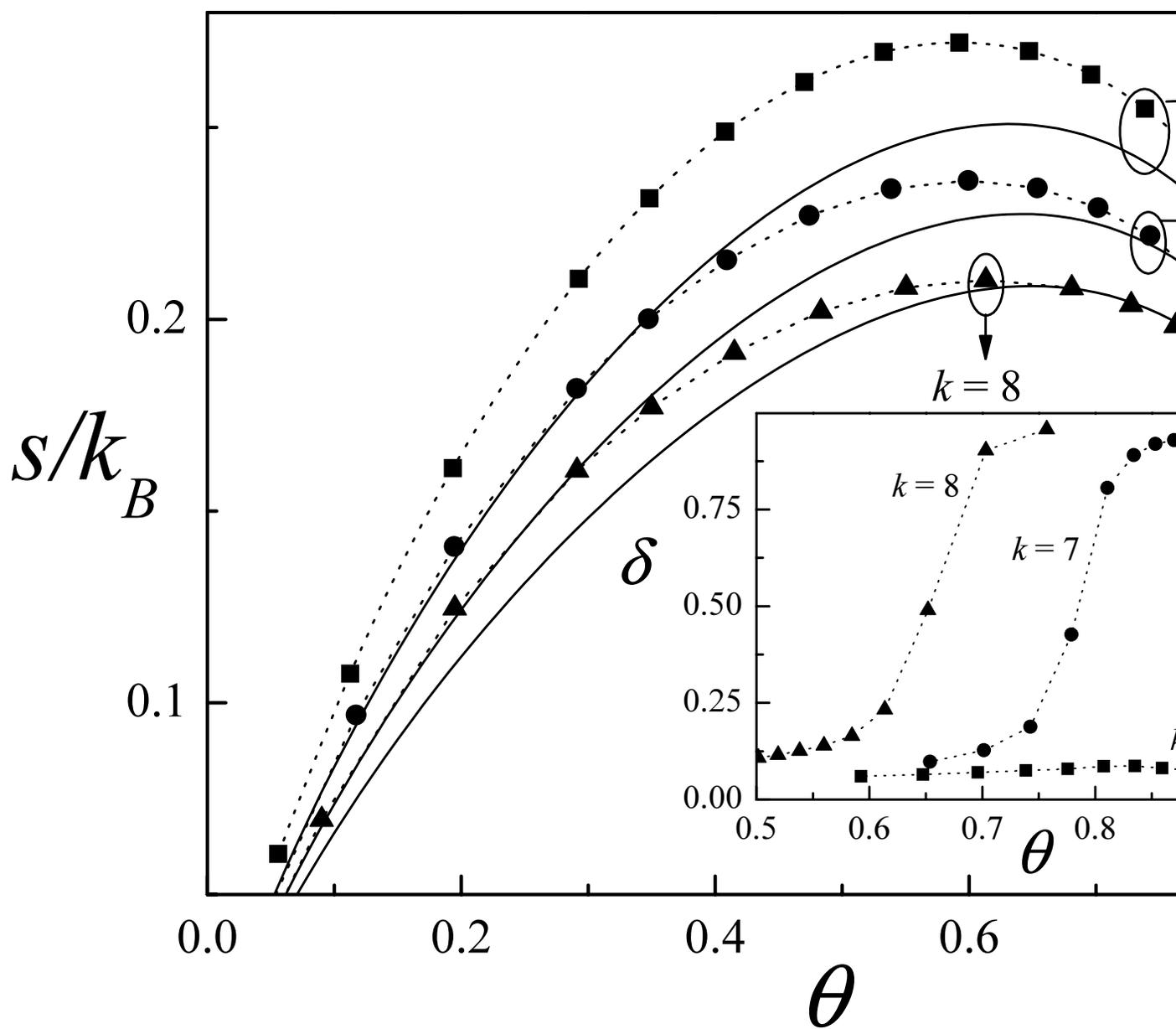
\noindent Fig. 1: Configurational entropy per site (in units of
$k_B$) versus surface coverage for different $k$ as indicated and
$L/k=20$. Dotted line and symbols represent MC results for
triangular lattices, and solid lines correspond to exact results
for one-dimensional systems. Inset: Surface coverage dependence of
the nematic order parameter for different $k$ as indicated and
$L/k=20$.

\noindent Fig. 2: (a) Curves of $U_L$ vs $\theta$ for square
lattices, $k=10$ and different lattice sizes as indicated. From
their intersections one obtains $\theta_c$. Inset: Size dependence
of the order parameter, $\delta$, as a function of coverage. (b)
Same as part (a) for triangular lattices.

\noindent Fig. 3: Simulated results for critical density
$\theta_c$ as a function of $k$. The symbology is indicated in the
figure.

\noindent Fig. 4: (a) An example of a ``small window" of side
$l=8$, which is embedded in an infinite lattice. Open squares
correspond to empty sites and black bars represent adsorbed
$k$-mers with $k=4$. (b) Same as part (a) for rescaled values of
$l$ and $k$. In this case, $l=24$ and $k=12$.

\noindent Fig. 5: Theoretical results for critical density
$\theta_c$ as a function of $k$. The symbology is indicated in the
figure.

\newpage


\end{large}


\begin{thebibliography}{99}
\bibitem{FOOT1} Note that the expression $\theta=kN/M$  is valid in the thermodynamic limit, where the
contribution to the local coverage due to $k$-mers adsorbed on the
window boundaries is negligible.
\end{thebibliography}
\end{document}